\documentclass{PoS}
\usepackage{subfig,amsmath,amssymb,cite,multirow}

\title{Signatures of Dark Star Remnants in the Galactic Halo}

\ShortTitle{Signatures of Dark Star Remnants in the Galactic Halo}

\author{\speaker{Pearl Sandick}\\
        Theory Group and Texas Cosmology Center, The University of Texas at Austin, TX 78712, USA\\
        E-mail: \email{pearl@physics.utexas.edu}}

\author{J\"{u}rg Diemand\\
        Institute for Theoretical Physics, University of Z\"urich, CH-8057, Switzerland\\
        E-mail: \email{diemand@physik.uzh.ch}}

\author{Katherine Freese\\
        Michigan Center for Theoretical Physics, University of Michigan, Ann Arbor, MI 48109\\
        E-mail: \email{ktfreese@umich.edu}}

\author{Douglas Spolyar\\
        Center for Particle Astrophysics, Fermi National Accelerator Laboratory, Batavia, IL  60510\\
        Department of Astronomy and Astrophysics, The University of Chicago, Chicago, IL 60637\\
        E-mail: \email{dspolyar@fnal.gov}}

\abstract{The very first stars likely formed from metal-free, molecular hydrogen-cooled gas at the center of dark matter minihalos.  Prior to nuclear fusion, these stars may have been supported by dark matter heating from annihilations in the star, in which case they could have grown to be quite massive before collapsing to black holes. Many remnant black holes and their surrounding dark matter density spikes may be part of our Milky Way halo today.  Here we explore the gamma-ray signatures of dark matter annihilations in the dark matter spikes surrounding these black holes for a range of star formation scenarios, black hole masses, and dark matter annihilation modes. Data from the Fermi Gamma-Ray Space Telescope are used to constrain models of dark matter annihilation and the formation of the first stars.}

\FullConference{Identification of Dark Matter 2010-IDM2010\\
		July 26-30, 2010\\
		Montpellier France}

\begin{document}

\section{Introduction}

The very first generation of stars, known as Population III.1, likely formed from the pristine gas at the centers of $\sim 10^6 M_\odot$ dark matter minihalos at $z \gtrsim 10$~\cite{HTL1996}.  The response of the dark matter in a minihalo to the formation of a compact baryonic object at its center is a contraction of the dark matter density profile in and around the object.  When the first stars, expected to be $\gtrsim 100 M_\odot$, ended their lives by collapsing to black holes\footnote{Stars in the mass range $\sim 140-260 M_\odot$ would have ended their lives as pair instability supernovae, leaving no remnants~\cite{hegerwoosley}.  We do not consider these objects here.}, each remnant remained surrounded by a region of enhanced dark matter density, which we call a {\it dark matter spike}. 
In fact, if the first stage of stellar evolution is a Dark Star (DS) phase, during which the star is powered by dark matter annihilations, the first stars would have grown to be even larger, leaving correspondingly larger black holes and surrounding dark matter spikes.

Many of these spikes, remnants of the formation of the first stars, may have merged into our Galactic halo, constituting Milky Way dark matter substructure today.  Here we report on the results of Ref.~\cite{dmspikes}: the gamma-ray flux from dark matter annihilations in the dark matter spikes in our Galactic halo, and how we can use data from the Fermi Gamma-Ray Space Telescope (FGST) to constrain models of Population III.1 star formation and/or dark matter annihilation.

\section{The First Stars and Their Dark Matter Spikes}

Population III.1 stars could only form when a cooling mechanism for the collapse of the baryonic cloud arose. The first accessible mechanism to cool the gas was via excitations of molecular hydrogen, the fraction of which present in a minihalo is related to the temperature, which in turn can be written in terms of the mass and redshift of the minihalo. We use the parametrization of Ref.~\cite{Hcooling} for the minimum halo mass in which star formation could occur:
\begin{equation}
M_{min}^{halo} \approx 1.54 \times 10^5 M_\odot \Big(\frac{1+z}{31}\Big)^{-2.074}.
\label{eq:Hcooling}
\end{equation}
We take the maximum halo mass for Population III.1 star formation to be $M_{max}^{halo}=10^7 M_\odot$, though our results are not sensitive to this choice. 

At some time between the beginning of Population III.1 star formation and the end of reionization at $z \sim 6$, massive Population III.1 (and DS) formation must have given way to subseqent formation of less-massive stars~\cite{greifbromm2006}, however there are few constraints on when this transition occured.  Here we consider three scenarios for the termination of Population III.1 star formation at redshift $z_f$.  These scenarios are hereafter noted as Early, Intermediate, and Late, following Ref.~\cite{greifbromm2006}, with $z_f \approx 23$, $15$, and $11$, respectively.
For each case, we assume that Population III.1 star formation was possible in any minihalo with a mass between $M_{min}^{halo}$ and $M_{max}^{halo}$ at redshift $z \geq z_f$.
 
In fact, not every minihalo meeting the above criteria must have hosted a Population III.1 star.  We therefore parametrize the fraction that did as $f_{DS}$. Neglecting black hole mergers, a discussion of which can be found in~\cite{dmspikes}, the comoving number density of black holes as a function of redshift is then
\begin{equation}
N_{BH}(z) = f_{DS} \, N_{halo}(z),
\end{equation}
where $N_{halo}(z)$ is the comoving number density of minihalos in which Population III.1 star formation was possible.

\begin{figure}[h]
\begin{center}
\includegraphics[width=.6\textwidth]{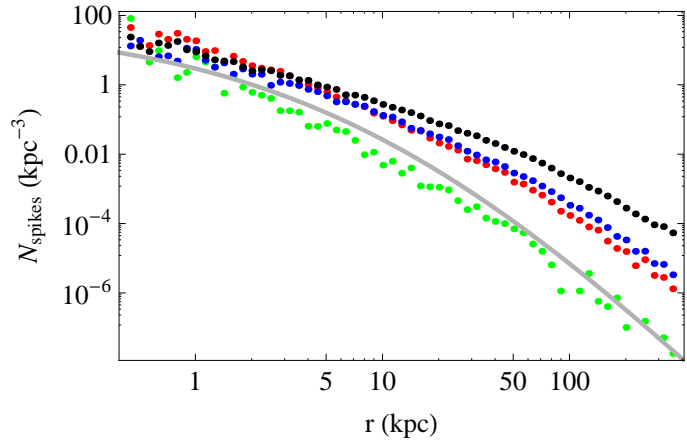}
\end{center}
\caption{The number density of black hole spikes in the Milky Way as a function of Galactic radius for star formation models with Early (green), Intermediate (red), and Late (blue) $z_f$ as described in the text.  
The black points illustrate the total dark matter density profile at $z=0$.  Also shown as a solid grey curve is the analytical fit found in Ref.~\cite{bzs,tabp}. 
\label{fig:spikedists}}
\end{figure}

Assuming that each viable minihalo hosted a Population III.1 star ($f_{DS}=1$), we determine the $z=0$ distribution of dark matter spikes throughout the Galactic halo from the Via Lactea II (VL-II) cosmological N-body simulation~\cite{vl2}. In Fig.~\ref{fig:spikedists} we show the number densities of dark matter spikes inside the Milky Way halo as functions of Galactic radius for Early, Intermediate, and Late $z_f$. In the Early case, Pop.~III.1 star formation terminates at $z\approx 23$, so there were the fewest stars, and therefore the fewest black holes and surviving density spikes today; 409 spikes in our Galactic halo. In the Intermediate and Late cases, we find 7983 and 12416 dark matter spikes in our Galactic halo, respectively. In a similar analysis, Ref.~\cite{bzs} found the distribution of spikes given by the grey curve in Fig.~\ref{fig:spikedists}, resulting in $\sim1027$ spikes in our Galactic halo.  For comparison, the total dark matter density profile at $z=0$ in VL-II is also shown; although the normalization of these points is
arbitrary, it is useful to illustrate that the total dark matter profile is more extended than the distribution of  black holes with dark matter spikes.

The density profile of an individual dark matter spike surrounding a black hole is determined by adiabatic contraction of the dark matter halo around the central mass.  We assume Navarro, Frenk, and White (NFW) profiles for both the baryons and dark matter, and compute the contracted dark matter profile using the Blumenthal {\it et al.} prescription for adiabatic contraction~\cite{blum}. The resulting halo profiles are roughly power-law in nature with a high density cut-off imposed to account for the annihilation of the dark matter in the spike; 
$\rho_{max} = m_{\chi}/(\langle \sigma v \rangle t_{BH})$,
where $m_\chi$ is the mass of the dark matter particle, $\langle \sigma v \rangle$ its annihilation cross section times velocity, and
$t_{BH}$ is the lifetime of the central mass, roughly $1.3 \times 10^{10}$ years for a star that formed at $z=15$. 

While standard Population III.1 stars are expected to have masses of $\sim 100 M_\odot$, if dark matter is
 capable of self-annihilating, then the first stars may have been powered by dark matter annihilations for some period of time prior to nuclear fusion.  This first phase of stellar evolution is known as the Dark Star (DS) phase~\cite{spolyar08} and may have lasted anywhere from a few hundred thousand years to millions, or even billions, of years.  During the DS phase, the star remains cool enough to continue to accrete baryonic matter, and may grow to $\sim 1000M_\odot$~\cite{Freese:2008wh} or even as large as $\gtrsim 10^5 M_\odot$~\cite{Freese:2010re}, depending on the details of how the dark matter is depleted and replenished in the star.

When the dark matter fuel inside a DS runs out, it collapses and heats up to become a standard fusion-powered star, which, at the end of its life, will likely undergo core collapse leaving a black hole remnant. We use the potential existence of the DS phase to motivate consideration of black holes as large as $10^5 M_\odot$.

%%%%%%%%%%%%%%%%%%%%%%%%%%%%%%%%%%%%%%%%%%%%%%%%%%%%%%%%%%%%%%%%%%%%%%

\section{Gamma Ray Signal from Dark Matter Annihilations}
\label{sec:signal}

For a Majorana dark matter particle with mass $m_\chi$ and annihilation cross section times velocity $\langle \sigma v \rangle$, the rate of WIMP annihilations in a dark matter spike is
\begin{equation}
\Gamma = \frac{\langle \sigma v \rangle}{2 m_\chi^2}\int_{r_{min}}^{r_{max}} dr \, 4 \pi r^2 \, \rho_{DM}^2(r),
\label{eq:rate}
\end{equation}
with $r_{min}$ and $r_{max}$ defining the volume of the dark matter spike in which annihilations occur.

We choose as a benchmark scenario $\langle \sigma v \rangle = 3 \times 10^{-26}$ cm$^3$s$^{-1}$, in agreement with the measured dark matter abundance today for thermal WIMP dark matter, and consider several WIMP candidates, defined by mass and annihilation channel. We consider annihilations of WIMPs with masses from 100 GeV to 2 TeV to Standard Model final states $b \bar{b}$, $W^+W^-$, $\tau^+ \tau^-$, and $\mu^+ \mu^-$. 
The resulting spectrum of photons from annihilation to final state $f$ is $dN_f/dE$.  For $\chi \chi \rightarrow \mu^+\mu^-$, the photon spectrum comes only from final state radiation. In the following analysis, we assume that the branching fraction to each final state is 1, though, in principle, some combination of final states is possible.

%%%%%%%%%%%%%%%%%%%%%%%%%%%%%%%%%%%%%%%%%%%%%%%%%%%%%%%%%%%%%%%%%%%%%%%%%%

The differential flux of neutral particles from annihilations to final state $f$ in a dark matter spike with radius $r_{max}$ located some distance $D$ from our Solar System is given by
\begin{equation}
\frac{d\Phi_f}{dE}=\frac{\Gamma}{4 \pi D^2} \frac{dN_f}{dE},
\end{equation}
for $D \gg r_{max}$.  
If a single spike is a sufficiently bright and compact source of gamma-rays, it may have been identified as a point source and recorded in the FGST First Source Catalog~\cite{fgstFSC}.
Spikes bright enough to be identified as point sources must not be brighter than the brightest source in the the FGST catalog.  Requiring that the flux not exceed this brightness establishes a minimal distance, $D_{min}^{PS}$, beyond which the spike must be located. Similarly, we can define a maximal distance, $D_{max}^{PS}$, as the distance beyond which a spike would not be bright enough to have been detected at $5 \sigma$ significance (see~\cite{dmspikes} for details). All spikes which could have been identified as point sources lie between $D_{min}^{PS}$ and $D_{max}^{PS}$.

\begin{figure}[h!]
\begin{center}
\subfloat{\includegraphics[width=0.45\textwidth]{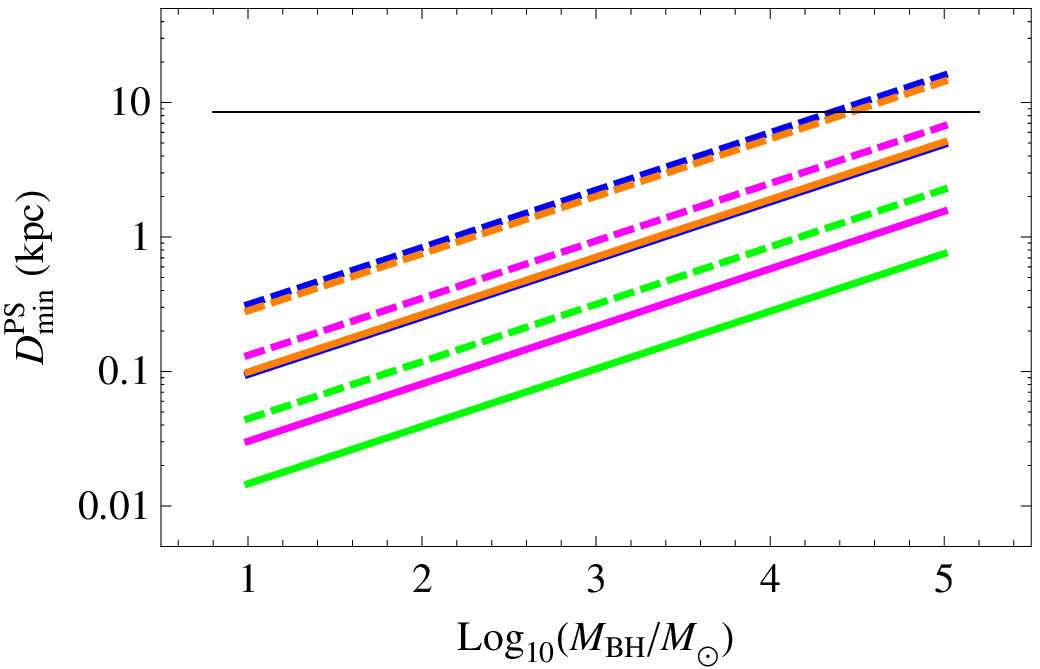}}\hspace{.3cm}
\subfloat{\includegraphics[width=0.45\textwidth]{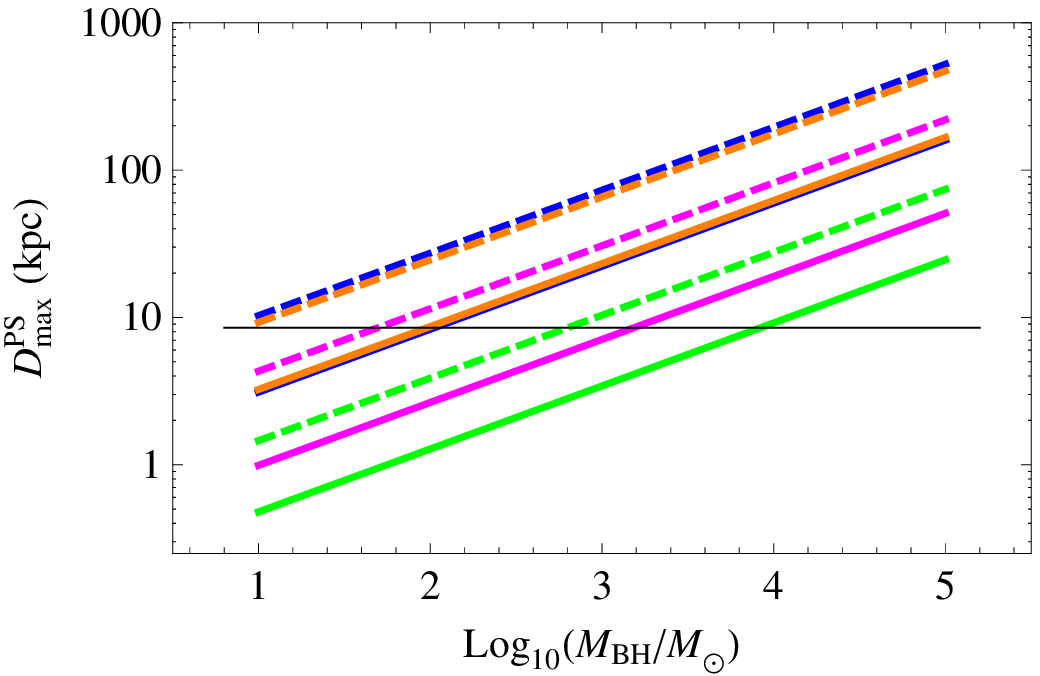}}
\caption{In the left panel, we display the minimal distance from our Solar System of a single spike such that it does not exceed the flux of the brightest source in the FGST First Source Catalog, $D_{min}^{PS}$, as a function of central black hole mass for Intermediate $z_f$.  From top to bottom, the contours are for 100 GeV WIMPs annihilating to $b\bar{b}$ (blue dashed), 100 GeV WIMPs annihilating to $W^+W^-$ (orange dashed), 100 GeV WIMPs annihilating to $\tau^+\tau^-$ (magenta dashed), 1 TeV WIMPs annihilating to $b\bar{b}$ (blue solid), 1 TeV WIMPs annihilating to $W^+W^-$ (orange solid), 100 GeV WIMPs annihilating to $\mu^+\mu^-$ (green dashed), 1 TeV WIMPs annihilating to $\tau^+\tau^-$ (magenta), and 1 TeV WIMPs annihilating to $\mu^+\mu^-$ (green solid).  In the right panel, we show the maximal distance from our Solar System of a single spike such that it would appear as a $\gtrsim 5\sigma$ point source to Fermi, $D_{max}^{PS}$, for the same cases. The horizontal black line indicates the distance to the Galactic center.
\label{fig:PointSourceDist}}
\end{center}
\end{figure}

In Fig.~\ref{fig:PointSourceDist}, we display in the left panel the minimal distance, $D_{min}^{PS}$, at which a point source may be located in order not to exceed the largest flux from any point source measured by FGST for eight example WIMP annihilation cases.  
For most of the dark matter models shown, if the central black hole is $\lesssim 10^3 M_\odot$, spikes may be located within 1 kpc of our Solar System. However, for all choices of $z_f$ considered here, we expect less than one dark matter spike within 1 kpc of our Solar System, even for $f_{DS}=1$.
In the right panel of Fig.~\ref{fig:PointSourceDist}, we show the maximal distance, $D_{max}^{PS}$, at which a single dark matter spike would have been identified by FGST as a $\gtrsim 5\sigma$ point source. For 100 GeV WIMPs annihilating to $b\bar{b}$ or $W^+W^-$ around large black holes, as in the upper right portion of the plot, all dark matter spikes in the Milky Way halo would have been identified as point sources by FGST.

%%%%%%%%%%%%%%%%%%%%%%%%%%%%%%%%%%%%%%%%%%%%%%%%%%%%%%%%%%

\begin{figure}[h!]
\begin{center}
\subfloat{\includegraphics[width=.45\textwidth]{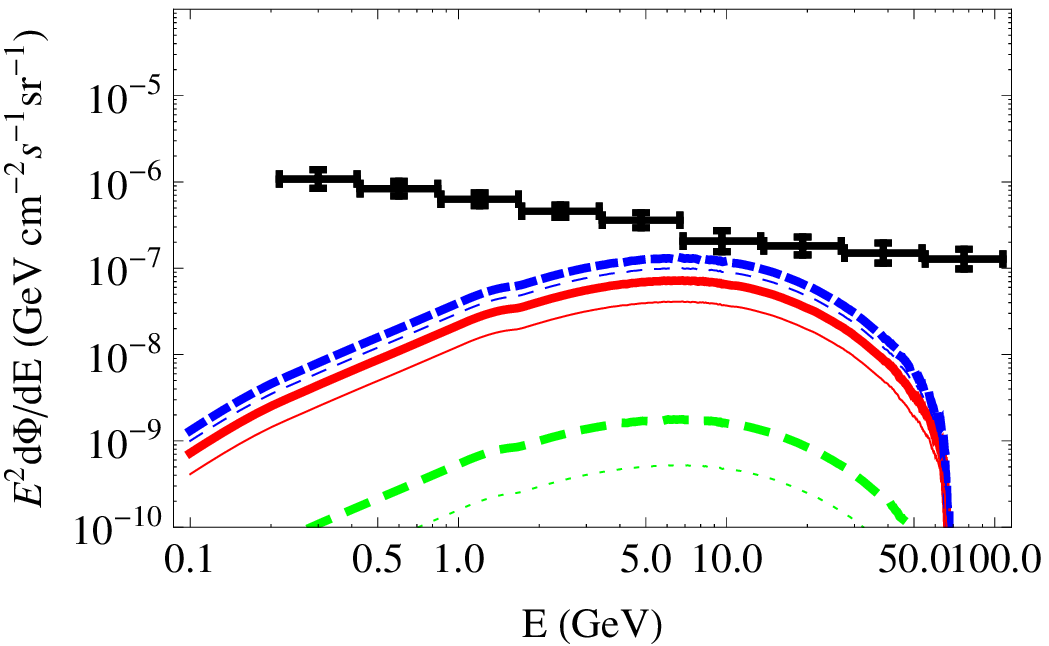}}\hspace{3mm}
\subfloat{\includegraphics[width=.45\textwidth]{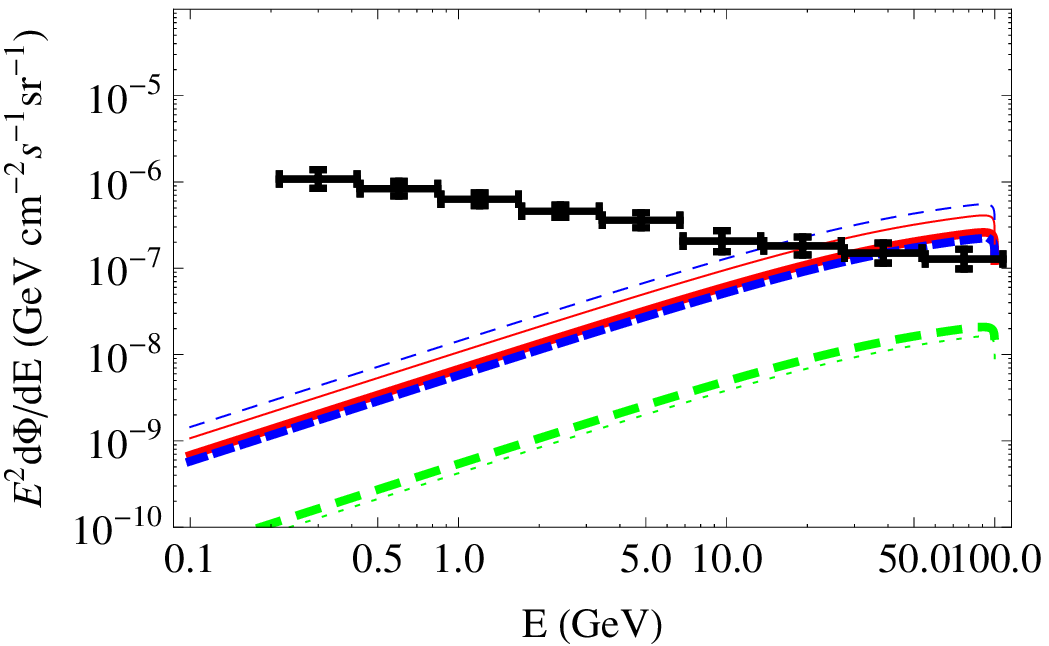}}
\end{center}
\caption{Diffuse gamma-ray flux from dark matter spikes for annihilations of 100 GeV WIMPs to $b\bar{b}$ (left) and $\mu^+\mu^-$ (right).  In each panel, Early (green, dotted), Intermediate (red, solid), and Late (blue, dashed) $z_f$ are shown for central black holes of mass $10^2 M_\odot$ (thick curves) and $10^3 M_\odot$ (thin curves). The black points represent the FGST-measured diffuse gamma-ray flux~\cite{fgstEGB}. 
\label{fig:Results}}
\end{figure}

Many of the dark matter spikes in the Milky Way halo may be too faint to be identified as point sources by FGST and would therefore contribute to the diffuse flux.
Based on the simulated distribution of Population III.1/DS remnant spikes in the Milky Way halo from VL-II, we calculate the contribution to the diffuse gamma ray flux from all spikes that are decidedly too faint to have been identified as point sources by FGST. These results are presented in Fig.~\ref{fig:Results} for annihilations of 100 GeV WIMPs to $b\bar{b}$ (left panel) and $\mu^+\mu^-$ (right panel) along with the FGST-measured diffuse gamma-ray flux~\cite{fgstEGB}. We see that the all-sky averaged diffuse flux is well below that measured by FGST in all models shown at low energies, but can rise to the level of the FGST-measured flux at intermediate and higher energies.

%%%%%%%%%%%%%%%%%%%%%%%%%%%%%%%%%%%%%%%%%%%%%%%%%

\section{Constraining $f_{DS}$}
\label{sec:fDS}

\begin{figure}[h!]
\centering
\subfloat{\includegraphics[width=.45\textwidth]{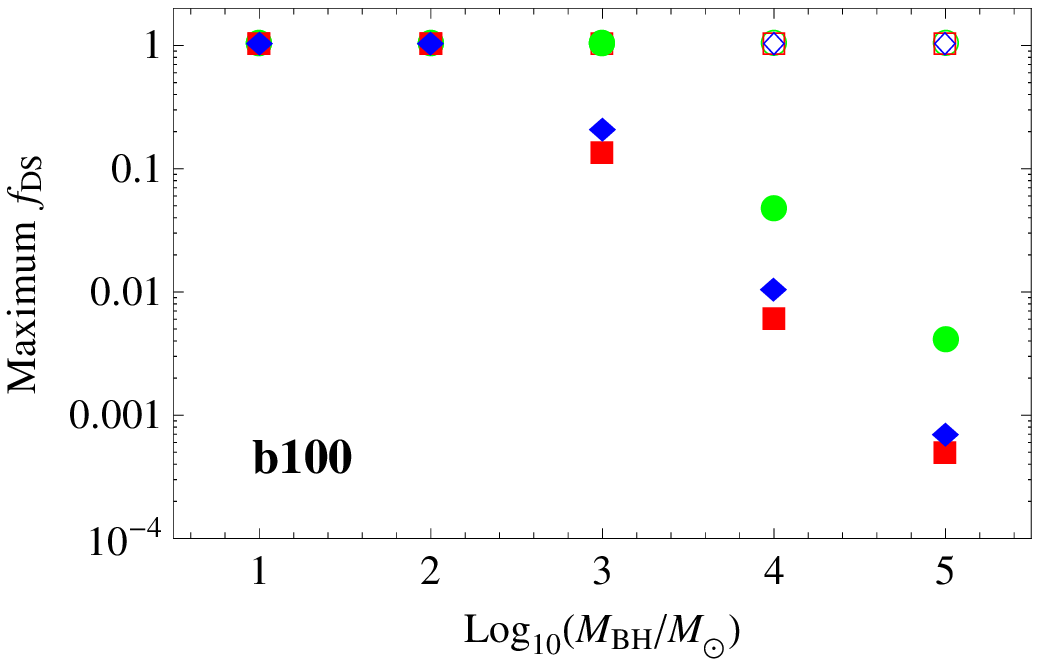}}\hspace{3mm}
\subfloat{\includegraphics[width=.45\textwidth]{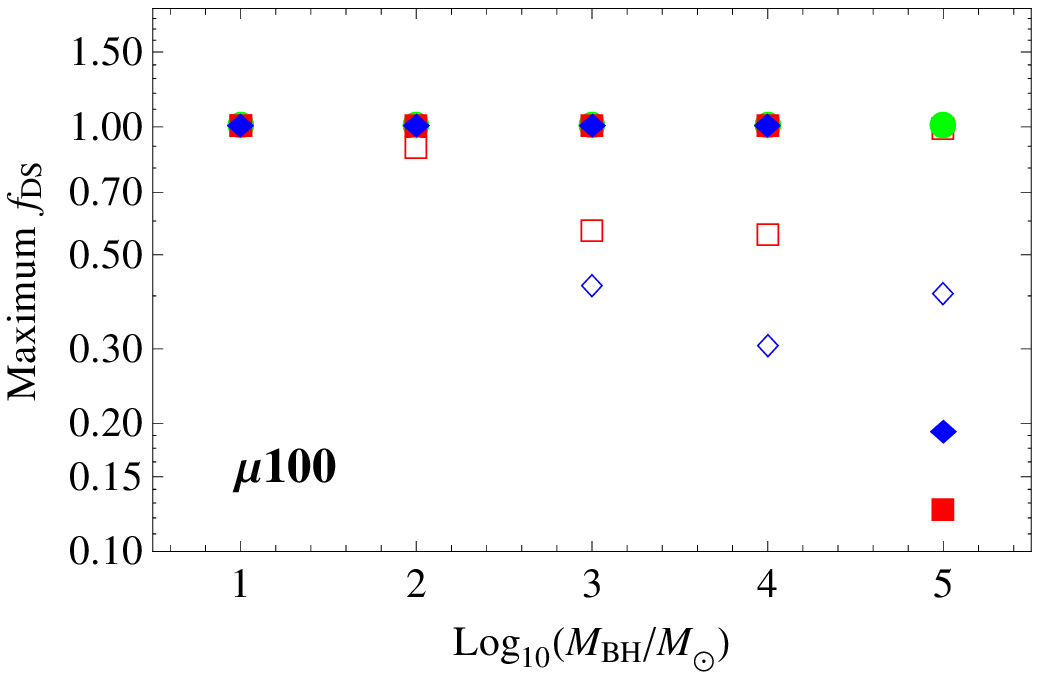}}
\caption{Maximum $f_{DS}$ as a function of central black hole mass for 100 GeV WIMPs annihilating to $b\bar{b}$ (left) and $\mu^+\mu^-$ (right). Green circles, red squares, and blue diamonds are for Early, Intermediate, and Late $z_f$, respectively. The solid markers are the limits from point source brightness, while the open markers are from the diffuse gamma-ray flux, as described in the text. 
\label{fig:fDSmax}}
\end{figure}

To this point we have focused only on $f_{DS}=1$, though in fact $f_{DS}$ may be $\ll 1$.  It is possible to constrain $f_{DS}$ in two ways:  First, from the FGST measurement of the diffuse gamma-ray flux, by requiring that the diffuse flux from dark matter annihilations around spikes in the Milky Way halo not exceed the measured flux in any of the nine FGST energy bins in Ref.~\cite{fgstEGB} by more than $3\sigma$.  The resulting maximal values of $f_{DS}$ are presented in Fig.~\ref{fig:fDSmax} for annihilations of 100 GeV WIMPs to $b\bar{b}$ (left panel) and $\mu^+\mu^-$ (right panel) as the open points, with Early, Intermediate, and Late $z_f$ represented in green, red, and blue, respectively.

A second way to constrain $f_{DS}$ is by requiring that one should expect to find less than one dark matter spike within $D_{min}^{PS}$, the minimal distance at which a spike may be located such that it is not brighter than the brightest point source in the FGST First Source Catalog~\cite{fgstFSC}. 
The maximum $f_{DS}$ as found in this method is displayed in Fig.~\ref{fig:fDSmax} as the solid points in each scenario.  These are the most conservative limits on $f_{DS}$, as they are based on the flux not exceeding that of the brightest FGST point source, known to be the Vela pulsar.  If we require that the flux from a dark matter spike not exceed that of the brightest unassociated point source, stronger limits on $f_{DS}$ would be obtained, though there is no gaurantee that there is not a dark matter spike along our line of sight to Vela. 

%%%%%%%%%%%%%%%%%%%%%%%%%%%%%%%%%%%%%

\section{Conclusions}

We have show how FGST gamma-ray observations can be used to place limits on the properties of dark matter
spikes around black holes in our Galactic halo.  We compare the gamma-ray flux from dark matter annihilations around spikes to both the gamma-ray flux from point sources and the diffuse
flux observed by FGST, and constrain the maximum fraction of viable minihalos that could have hosted Population III.1/DS star formation. In general, it is clear that the
bounds are the strongest for the largest black hole masses and if star formation persisted to low redshift
(in which case the total possible number of dark matter spikes in our Galactic halo is largest). These limits also depend sensitively on the dark matter annihilation channel, as one can see by comparing the results for annihilations to $b\bar{b}$ and $\mu^+\mu^-$ final states. If/when
the annihilation properties of particle dark matter become known, we may be provided with some
interesting hints about the formation of the first stars.

%%%%%%%%%%%%%%%%%%%%%%%%%%%%%%%%%%%%%%%%%%%

\section*{Acknowledgements}
\noindent P.S.~is supported by the NSF under Grant Numbers PHY-0969020 and PHY-0455649. J.D.~is supported by the Swiss National Science Foundation. K.F.~is supported by the DOE and the MCTP. D.S.~is supported by the DOE. We would also like to thank the organizers of IDM2010.


\begin{thebibliography}{99}

\bibitem{HTL1996}
  Z.~Haiman, A.~A.~Thoul and A.~Loeb,
  %``Cosmological formation of low mass objects,''
  Astrophys.\ J.\  {\bf 464}, 523 (1996)
  [arXiv:astro-ph/9507111].
  %%CITATION = ASJOA,464,523;%%
  
\bibitem{hegerwoosley}
  A.~Heger and S.~E.~Woosley,
  %``The Nucleosynthetic Signature of Population III,''
  Astrophys.\ J.\  {\bf 567}, 532 (2002)
  [arXiv:astro-ph/0107037].
  %%CITATION = ASJOA,567,532;%%

 \bibitem{dmspikes}
  P.~Sandick, J.~Diemand, K.~Freese and D.~Spolyar,
  %``Black Holes in our Galactic Halo: Compatibility with FGST and PAMELA Data
  %and Constraints on the First Stars,''
  arXiv:1008.3552 [astro-ph.CO].
  %%CITATION = ARXIV:1008.3552;%%

\bibitem{Hcooling}
  M.~Trenti and M.~Stiavelli,
  %``The Formation Rates of Population III Stars and Chemical Enrichment of
  %Halos during the Reionization Era,''
  Astrophys.\ J.\  {\bf 694}, 879 (2009)
  [arXiv:0901.0711 [astro-ph.CO]].
  %%CITATION = ASJOA,694,879;%%

\bibitem{greifbromm2006}
  T.~H.~Greif and V.~Bromm,
  %``Two populations of metal-free stars in the early Universe,''
  Mon.\ Not.\ Roy.\ Astron.\ Soc.\  {\bf 373}, 128 (2006)
  [arXiv:astro-ph/0604367].
  %%CITATION = MNRAA,373,128;%%

\bibitem{vl2}
  J.~Diemand, M.~Kuhlen, P.~Madau, M.~Zemp, B.~Moore, D.~Potter and J.~Stadel,
  %``Clumps and streams in the local dark matter distribution,''
  Nature {\bf 454}, 735 (2008)
  [arXiv:0805.1244 [astro-ph]].
  %%CITATION = NATUA,454,735;%%

\bibitem{bzs}
  G.~Bertone, A.~R.~Zentner and J.~Silk,
  %``A new signature of dark matter annihilations: Gamma-rays from
  %intermediate-mass black holes,''
  Phys.\ Rev.\  D {\bf 72}, 103517 (2005)
  [arXiv:astro-ph/0509565].
  %%CITATION = PHRVA,D72,103517;%%

\bibitem{tabp}
  M.~Taoso, S.~Ando, G.~Bertone and S.~Profumo,
  %``Angular correlations in the cosmic gamma-ray background from dark matter
  %annihilation around intermediate-mass black holes,''
  Phys.\ Rev.\  D {\bf 79}, 043521 (2009)
  [arXiv:0811.4493 [astro-ph]].
  %%CITATION = PHRVA,D79,043521;%%

\bibitem{blum}
  G.~R.~Blumenthal, S.~M.~Faber, R.~Flores and J.~R.~Primack,
  %``Contraction Of Dark Matter Galactic Halos Due To Baryonic Infall,''
  Astrophys.\ J.\  {\bf 301}, 27 (1986).
  %%CITATION = ASJOA,301,27;%%

\bibitem{spolyar08}
  D.~Spolyar, K.~Freese and P.~Gondolo,
  %``Dark matter and the first stars: a new phase of stellar evolution,''
  Phys.\ Rev.\ Lett.\  {\bf 100}, 051101 (2008)
  [arXiv:0705.0521 [astro-ph]].

\bibitem{Freese:2008wh}
  K.~Freese, P.~Bodenheimer, D.~Spolyar and P.~Gondolo,
  %``Stellar Structure of Dark Stars: a first phase of Stellar Evolution due to
  %Dark Matter Annihilation,''
  Astrophys.\ J.\  {\bf 685}, L101 (2008)
  [arXiv:0806.0617 [astro-ph]].
  %%CITATION = ASJOA,685,L101;%%

\bibitem{Freese:2010re}
  K.~Freese, C.~Ilie, D.~Spolyar, M.~Valluri and P.~Bodenheimer,
  %``Supermassive Dark Stars: Detectable in JWST,''
  arXiv:1002.2233 [astro-ph.CO].
  %%CITATION = ARXIV:1002.2233;%%

\bibitem{fgstFSC}
  T.~L.~Collaboration,
  %``Fermi Large Area Telescope First Source Catalog,''
  arXiv:1002.2280 [astro-ph.HE].
  %%CITATION = ARXIV:1002.2280;%

\bibitem{fgstEGB}
  A.~A.~Abdo {\it et al.}  [The Fermi-LAT collaboration],
  %``The Spectrum of the Isotropic Diffuse Gamma-Ray Emission Derived From
  %First-Year Fermi Large Area Telescope Data,''
  Phys.\ Rev.\ Lett.\  {\bf 104}, 101101 (2010)
  [arXiv:1002.3603 [astro-ph.HE]].
  %%CITATION = PRLTA,104,101101;%%

\end{thebibliography}
\end{document}